\documentclass{pasj00}

\def\commenta{$^*$}
\def\commentb{$^\dagger$}
\def\commentc{$^\ddagger$}
\def\inpress{in press}
\def\arxiv#1{ (arXiv astro-ph/#1)}

\DeclareAbbreviation\JBAA{J. British Astron. Assoc.}
\DeclareAbbreviation\ibvs{Inf. Bull. Variable Stars}
\DeclareAbbreviation\SPIE{SPIE Proc.}

\def\ASPConf#1#2{ASP Conf. Ser. #1, #2}
\def\PublisherReidel{Dordrecht: D. Reidel Publishing Company}
\def\PublisherASP{San Francisco: ASP}

\newcounter{author}
\def\authorcount#1#2{\refstepcounter{author}\label{#1}
                     \altaffiltext{\ref{#1}}{#2}}

\begin{document}
\SetRunningHead{T. Kato et al.}{SDSS J080434.20$+$510349.2}

\Received{200X/XX/XX}
\Accepted{200X/XX/XX}

\title{SDSS J080434.20$+$510349.2: Eclipsing WZ Sge-Type Dwarf Nova with Multiple Rebrightenings}

\author{Taichi~\textsc{Kato},\altaffilmark{\ref{affil:Kyoto}*}
        Elena~P.~\textsc{Pavlenko},\altaffilmark{\ref{affil:Pavlenko}}
        Hiroyuki~\textsc{Maehara},\altaffilmark{\ref{affil:HidaKwasan}}
        Kazuhiro~\textsc{Nakajima},\altaffilmark{\ref{affil:Njh}}
        Maksim~\textsc{Andreev},\altaffilmark{\ref{affil:Terskol}} \\
        Sergei~Yu.~\textsc{Shugarov},\altaffilmark{\ref{affil:Shugarov}}$^,$\altaffilmark{\ref{affil:Shugarov2}}
        Pierre~\textsc{de Ponthi\`ere},\altaffilmark{\ref{affil:Ponthiere}}
        Steve~\textsc{Brady},\altaffilmark{\ref{affil:Brady}}
        Geir~\textsc{Klingenberg},\altaffilmark{\ref{affil:Klingenberg}}
        Jeremy~\textsc{Shears},\altaffilmark{\ref{affil:Shears}} \\
        Akira~\textsc{Imada},\altaffilmark{\ref{affil:Imada}}
}

\authorcount{affil:Kyoto}{
     Department of Astronomy, Kyoto University, Kyoto 606-8502}
\email{$^*$tkato@kusastro.kyoto-u.ac.jp}

\authorcount{affil:Pavlenko}{
     Crimean Astrophysical Observatory, 98409, Nauchny, Crimea, Ukraine}

\authorcount{affil:HidaKwasan}{
     Kwasan and Hida Observatories, Kyoto University, Yamashina,
     Kyoto 607-8471}

\authorcount{affil:Njh}{
     Variable Star Observers League in Japan (VSOLJ),
     124 Isatotyo, Teradani, Kumano, Mie 519-4673}

\authorcount{affil:Terskol}{
     Institute of Astronomy, Russian Academy of Sciences, 361605 Peak Terskol, Kabardino-Balkaria, Russia}

\authorcount{affil:Shugarov}{
     Sternberg Astronomical Institute, Moscow University, Universitetsky
     Ave., 13, Moscow 119992, Russia}

\authorcount{affil:Shugarov2}{
     Astronomical Institute of the Slovak Academy of Sciences, 05960,
     Tatranska Lomnica, the Slovak Republic}

\authorcount{affil:Ponthiere}{
     American Association of Variable Star Observers (AAVSO),
     15 rue Pr\'e Mathy, 5170 Lesve (Profondeville), Belgium}

\authorcount{affil:Brady}{
     5 Melba Drive, Hudson, NH 03051, USA}

\authorcount{affil:Klingenberg}{
     Variable Star Section, Norwegian Astronomical Society, PO Box 1029 Blindern, 0315 Oslo, Norway}

\authorcount{affil:Shears}{
     ``Pemberton'', School Lane, Bunbury, Tarporley, Cheshire, CW6 9NR, UK}

\authorcount{affil:Imada}{
     Department of Physics, Faculty of Science, Kagoshima University,
     1-21-35, Kohrimoto, Kagoshima, 890-0065}



\maketitle

\begin{abstract}
   We observed the 2006 superoutburst of SDSS J080434.20$+$510349.2
during its plateau phase, rebrightening phase, and post-superoutburst
final decline.  We found that this object is a grazing eclipsing
system with a period of 0.0590048(2) d.  Well-defined eclipses were
only observed during the late stage of the superoutburst plateau and
the depth decreased during the subsequent stages.
We determined the superhump period during the superoutburst plateau
to be 0.059539(11) d, giving a fractional superhump excess of 0.90(2) \%.
During the rebrightening and post-superoutburst phases, persisting
superhumps with periods longer than those of superhumps during the
plateau phase: 0.059632(6) during the rebrightening phase and 0.05969(4) d
during the final fading.  This phenomenon is very well in line with
the previously known long-period ``late superhumps'' in GW Lib, V455 And
and WZ Sge.
The amplitudes of orbital humps between different states of rebrightenings
suggest that these humps do not arise from the classical hot spot,
but are more likely a result of projection effect in a high-inclination
system.  There was no clear evidence for the enhanced hot spot during
the rebrightening phase.  We also studied previously reported
``mini-outbursts'' in the quiescent state and found evidence that
superhumps were transiently excited during these mini-outbursts.
The presence of grazing eclipses and distinct multiple rebrightenings
in SDSS J080434.20$+$510349.2 would provide a unique opportunity to
understanding the mechanism of rebrightenings in WZ Sge-type dwarf novae.
\end{abstract}

\section{Introduction}

   Dwarf novae (DNe) are a class of cataclysmic variables (CVs), which are
close binary systems consisting of a white dwarf and a red-dwarf secondary
transferring matter via the Roche-lobe overflow.
SU UMa-type dwarf novae are a class of DNe exhibiting superhumps during
their long, bright outbursts (superoutbursts), which is believed to
be a result of tidal instabilities caused by the 3:1 orbital resonance
in the accretion disk
[see e.g. \citet{vog80suumastars}; \citet{war85suuma} for basic observational
properties; see \citet{osa96review} for a theoretical review].

   WZ Sge-type dwarf novae (see e.g. \cite{bai79wzsge}; \cite{dow90wxcet};
\cite{kat01hvvir}) are a subgroup of dwarf novae characterized by
large-amplitude (typically $\sim$ 8 mag) superoutbursts with long
(typically $\sim$ 10 yr) recurrence times.  Although it has been proposed
that the properties of outbursts in WZ Sge-type dwarf novae can be
basically understood within the framework of the thermal-tidal
disk-instability model (see e.g. \cite{osa89suuma}) without requiring
an enhanced mass-transfer (\cite{osa95wzsge}; \cite{osa03DNoutburst}),
the existence of an enhanced mass-transfer still remains controversial
(cf. \cite{pat02wzsge}; \cite{ste01wzsgesecondary}; \cite{ham97wzsgemodel}).

   WZ Sge-type dwarf novae are known to show several unusual properties
during outburst, which are rarely seen in ordinary SU UMa-type dwarf novae.
They include double-wave early superhumps with periods close to the orbital
periods during the early stage of superoutbursts \citep{kat02wzsgeESH},
which are considered to be a result of 2:1 resonance \citep{osa02wzsgehump}.
The frequent existence of post-superoutburst rebrightenings is also
characteristic to WZ Sge-type dwarf novae.  \citet{ima06tss0222}
classified these rebrightening by their morphology (see also the
activity sequence in \cite{kat04egcnc}).  Type-A superoutbursts
(long-lasting distinct plateau after the termination of the main
superoutburst) and type-B superoutbursts (superoutburst followed by distinct
multiple rebrightenings) are unique to WZ Sge-type dwarf novae.
Up to now, several objects are known to have shown type-B superoutbursts
at least once: UZ Boo, EG Cnc (\cite{pat98egcnc}; \cite{kat04egcnc}),
AL Com \citep{uem08alcom}, 1RXS J023238.8$-$371812, ASAS 153616$-$0839.1,
OT J074727.6$+$065050 (for a general review on these objects,
see Kato et al. in preparation).  The mechanism of these rebrightening
is still in dispute: \citet{pat98egcnc} suggested that these rebrightenings
are produced by an enhanced mass-transfer following the superoutburst
while \citet{osa97egcnc}, \citet{osa01egcnc} proposed that they can be
reproduced if the viscosity in the accretion disk remained higher
after the termination of the superoutburst.  A search for
the observational evidence for the existence of an enhanced mass-transfer
is a key in discriminating these possibilities.

   WZ Sge-type dwarf novae have also been shown to exhibit long-lasting
superhumps whose periods are longer than the those of superhumps
during the superoutburst plateau.  \citet{kat08wzsgelateSH} argued
that these superhumps arise from the matter outside the 3:1 resonance,
and suggested that the stability of their periods can be understood
if the outer edge of the accretion disk is limited by the tidal
truncation radius.

   SDSS J080434.20$+$510349.2 (hereafter SDSS J0804) was discovered by
\citet{szk06SDSSCV5}.  The first-ever recorded, 2006 outburst was
detected by E. Pavlenko (vsnet-alert 8874).
\citet{pav07j0804} and \citet{she07j0804}
reported the detection of superhumps and discussed the WZ Sge-type
nature of this object.  The object is renowned for its eleven
post-superoutburst rebrightenings (\cite{pav07j0804}; for the
details of the light curve of the rebrightenings,
see e.g. \cite{pav09j0804}; subsection \ref{sec:reb} in this paper),
surpassing the record of six rebrightenings in EG Cnc.
\citet{zha08j0804} reported the detection of two ``mini-outbursts''
one year after the 2006 superoutburst, and reported double-humped
quiescent light curve having a period of 0.05900 d.
\citet{pav08j0804WD} reported that a 12.6-m periodicity, which can be
attributed to pulsations of the white dwarf, emerged during the interval
2006--2008.

\section{Observation and Data Analysis}

   The observations are composed of those obtained at
Crimean Astrophysical Observatory (CrAO in table \ref{tab:log}) using
the 2.6-m telescope and FLI 1001E CCD
(BJD 2453799, March 4 and BJD 2453856--2453857, Apr 30--May 1),
the 60-cm telescope and an Ap47p CCD (BJD 2453812--2453813, March 17 and 18)
during rebrightening phase,
the 38-cm telescope and an ST-7 CCD during the final fading
(BJD 2453816--2453828, April 21--May 13, except Apr 30 and May 1), 
and observations during the rebrightening and post-superoutburst phase:
Maehara (Mhh in table \ref{tab:log}) using a 25-cm telescope and
an ST-7 XME CCD camera and Nakajima (Njh) using a 25-cm telescope and
a CV-04 CCD camera, Shugarov (Shu) using a 70-cm telescope and an Ap47p CCD
at Sternberg Astronomical Institute (SAI), Moscow, and at Terskol, Caucasus
using a 60-cm telescope and an S2C CCD.
The details of CrAO, SAI and Terskol observations will be given in
Pavlenko et al., in preparation.
Maehara and Nakajima used the common comparison
star of TYC2 3414.1011.1 ($V=11.30$, $B-V=+0.44$) and performed aperture
photometry with IRAF\footnote{
  IRAF is distributed by the National Optical Astronomy Observatories for
  Research in Astronomy Inc. under cooperative agreement with the National
  Science Foundation.
} and FitsPhot 4.1,\footnote{
  FITS Photo is aperture photometry software developed by Kazuo Nagai.
  This software is available at
  $<$http://www.geocities.jp/nagai\_kazuo/index-e.html$>$.
} respectively, after standard flat-fielding and dark subtractions.
The exposure times of both observers were 30 s.
S. Brady (BXS) used a 40-cm telescope and an ST8XME CCD.  The exposure
time was 180 s.

   We also incorporated observations by P. de Ponthi\`ere (DPP),
G. Klingenberg (GK) and J. Shears (JSh)  published in
\citet{she07j0804}.
The times of observations were converted to Barycentric Julian Dates
(BJD) before the analysis.

   In measuring the times of (super)humps, we first corrected for
systematic differences between observers, and then subtracted the general
trend by fitting low-order (typically three to five) polynomials
for the superoutburst plateau and the final fading phase.
For the complex rebrightening phase, we subdivided the light curve
into segments of $\sim$1-d duration, and subtracted the trend by fitting
third order polynomials to individual segments that have sufficient
numbers of data points.

   We measured the times of superhump maxima by numerically
fitting a template superhump light curve around the times of observed
maxima.  This fitting methods have been proven to give three times
the precision compared to eye estimates
We did not use the full superhump cycle but used
phases $-$0.3 to 0.3 in order to pick the features of the hump maxima.
We used a phase-averaged (and spline-interpolated) mean light curve of
superhumps of GW Lib during the 2007 superoutburst (Imada et al.,
in preparation; \cite{kat08wzsgelateSH}) as the template,
which is one of the best-sampled objects among all SU UMa-type dwarf novae
and has the least scatter.  Although the actual superhumps in SDSS J0804
may be slightly different from the superhump profile of GW Lib, this
difference has been confirmed insignificantly to affect the period analysis.

   The summary of observations, with mean magnitudes, are listed in
table \ref{tab:log}.

\begin{table*}
\caption{Log of observations.}\label{tab:log}
\begin{center}
\begin{tabular}{ccccccc}
\hline\hline
Start\commenta & End\commenta & Mean Mag. & Error & $N$\commentb & Observer/Site & Filter\commentc \\
\hline
53799.3370 & 53799.4855 & 13.089 & 0.003 & 387 & CrAO & V \\
53800.3022 & 53800.6019 & 13.157 & 0.002 & 714 & DPP & C \\
53800.3022 & 53800.6936 & 13.226 & 0.003 & 262 & GK & C \\
53800.3270 & 53800.4053 & 13.180 & 0.004 & 100 & JSh & C \\
53800.5040 & 53800.7613 & 13.487 & 0.004 & 80 & BXS & C \\
53800.7085 & 53800.7109 & 13.385 & 0.021 & 2 & BXS & C \\
53801.3282 & 53801.6449 & 13.506 & 0.004 & 547 & DPP & C \\
53801.5080 & 53801.7985 & 13.970 & 0.010 & 136 & BXS & C \\
53801.9097 & 53801.9669 & 13.944 & 0.008 & 105 & Njh & C \\
53802.0926 & 53802.2587 & 14.237 & 0.006 & 405 & Mhh & C \\
53802.5078 & 53802.5241 & 15.064 & 0.016 & 10 & BXS & C \\
53803.0274 & 53803.0671 & 15.012 & 0.025 & 29 & Njh & C \\
53804.1999 & 53804.2779 & 15.266 & 0.021 & 8 & Shu & R \\
53805.1897 & 53805.1897 & 15.337 & -- & 1 & Shu & R \\
53806.5063 & 53806.5063 & 13.88 & -- & 1 & BXS & C \\
53808.1013 & 53808.1677 & 14.683 & 0.007 & 168 & Mhh & C \\
53809.0442 & 53809.2232 & 15.026 & 0.005 & 427 & Mhh & C \\
53810.0594 & 53810.2185 & 13.491 & 0.002 & 504 & Mhh & C \\
53811.5575 & 53811.5575 & 15.10 & -- & 1 & BXS & C \\
53812.1013 & 53812.1722 & 14.197 & 0.013 & 186 & Mhh & C \\
53812.2004 & 53812.4951 & 13.637 & 0.012 & 65 & CrAO & R \\
53812.5107 & 53812.5107 & 13.77 & -- & 1 & BXS & C \\
53813.2053 & 53813.2489 & 14.205 & 0.023 & 9 & CrAO & R \\
53813.9362 & 53814.1537 & 14.720 & 0.003 & 544 & Mhh & C \\
53814.9190 & 53815.1756 & 13.481 & 0.003 & 464 & Njh & C \\
53815.1552 & 53815.2421 & 13.436 & 0.002 & 301 & Mhh & C \\
53815.5159 & 53815.7091 & 14.048 & 0.011 & 40 & BXS & C \\
53816.0380 & 53816.2360 & 14.157 & 0.005 & 469 & Mhh & C \\
53816.0967 & 53816.2028 & 14.207 & 0.008 & 99 & Njh & C \\
53816.2489 & 53816.4769 & 14.609 & 0.008 & 98 & CrAO & R \\
53817.2745 & 53817.3938 & 14.725 & 0.014 & 55 & CrAO & R \\
53817.9192 & 53818.1567 & 13.687 & 0.005 & 275 & Njh & C \\
53818.9198 & 53819.1555 & 14.719 & 0.006 & 425 & Njh & C \\
53819.4225 & 53819.4487 & 15.039 & 0.014 & 12 & CrAO & R \\
53819.9167 & 53820.1485 & 13.471 & 0.003 & 396 & Njh & C \\
53821.2211 & 53821.4482 & 14.650 & 0.006 & 106 & CrAO & R \\
53821.6416 & 53821.6416 & 15.14 & -- & 1 & BXS & C \\
53822.2184 & 53822.3824 & 14.868 & 0.011 & 77 & CrAO & R \\
53823.2288 & 53823.4449 & 14.262 & 0.009 & 94 & CrAO & R \\
53823.9468 & 53824.1574 & 14.786 & 0.008 & 341 & Njh & C \\
53823.9987 & 53824.1651 & 14.727 & 0.005 & 420 & Mhh & C \\
53824.3207 & 53824.3207 & 14.945 & -- & 1 & Terskol & R \\
53824.7377 & 53824.7377 & 15.16 & -- & 1 & BXS & C \\
53824.9737 & 53824.9960 & 13.512 & 0.015 & 25 & Njh & C \\
53825.1800 & 53825.2558 & 13.506 & 0.004 & 183 & Mhh & C \\
53825.9262 & 53826.1050 & 14.203 & 0.005 & 325 & Njh & C \\
53826.0139 & 53826.2084 & 14.252 & 0.008 & 323 & Mhh & C \\
53826.2293 & 53826.4553 & 14.776 & 0.008 & 104 & Terskol & R \\
53827.2101 & 53827.2689 & 13.877 & 0.003 & 70 & Terskol & R \\
53827.9254 & 53828.0650 & 14.091 & 0.012 & 36 & Njh & C \\
53828.0679 & 53828.2160 & 14.245 & 0.005 & 385 & Mhh & C \\
\hline
  \multicolumn{7}{l}{\commenta BJD$-$2400000.} \\
  \multicolumn{7}{l}{\commentb Number of observations.} \\
  \multicolumn{7}{l}{\commentc C indicates unfiltered observations.} \\
\end{tabular}
\end{center}
\end{table*}

\addtocounter{table}{-1}
\begin{table*}
\caption{Log of observations.}
\begin{center}
\begin{tabular}{ccccccc}
\hline\hline
Start & End & Mean Mag. & Error & $N$ & Observer/Site & Filter \\
\hline
53828.2510 & 53828.4373 & 14.684 & 0.008 & 83 & CrAO & R \\
53828.5254 & 53828.5254 & 14.95 & -- & 1 & BXS & C \\
53828.9242 & 53828.9905 & 15.111 & 0.012 & 119 & Njh & C \\
53829.2372 & 53829.4310 & 15.162 & 0.011 & 91 & CrAO & R \\
53830.2264 & 53830.4592 & 14.228 & 0.008 & 109 & CrAO & R \\
53830.9355 & 53830.9553 & 14.610 & 0.031 & 14 & Njh & C \\
53831.2388 & 53831.5057 & 15.289 & 0.017 & 121 & CrAO & R \\
53832.2429 & 53832.3537 & 15.773 & 0.007 & 148 & Terskol & R \\
53832.9323 & 53833.1366 & 13.783 & 0.004 & 370 & Njh & C \\
53833.9342 & 53834.1025 & 14.938 & 0.008 & 535 & Mhh & C \\
53836.2744 & 53836.3005 & 15.315 & 0.057 & 12 & CrAO & R \\
53836.7084 & 53836.7084 & 15.93 & -- & 1 & BXS & C \\
53839.0047 & 53839.0103 & 15.813 & 0.030 & 8 & Njh & C \\
53839.2298 & 53839.4737 & 15.678 & 0.025 & 96 & CrAO & R \\
53841.5345 & 53841.5345 & 16.00 & -- & 1 & BXS & C \\
53842.3025 & 53842.3111 & 15.718 & 0.053 & 5 & CrAO & R \\
53845.5435 & 53845.5435 & 16.495 & 0.292 & 2 & BXS & C \\
53847.2571 & 53847.2878 & 15.952 & 0.032 & 14 & CrAO & R \\
53848.2584 & 53848.4358 & 15.975 & 0.011 & 82 & CrAO & R \\
53849.2920 & 53849.3139 & 16.050 & 0.015 & 11 & CrAO & R \\
53850.2500 & 53850.4200 & 16.104 & 0.012 & 78 & CrAO & R \\
53851.2624 & 53851.2798 & 16.119 & 0.090 & 8 & CrAO & R \\
53852.5674 & 53852.5674 & 16.55 & -- & 1 & BXS & C \\
53852.5888 & 53852.5888 & 16.36 & -- & 1 & BXS & C \\
53853.2874 & 53853.4418 & 16.151 & 0.017 & 63 & CrAO & R \\
53854.2914 & 53854.4387 & 16.198 & 0.013 & 68 & CrAO & R \\
53855.2543 & 53855.4105 & 16.211 & 0.014 & 74 & CrAO & R \\
53856.2632 & 53856.4095 & 16.223 & 0.009 & 112 & CrAO & V \\
53856.5491 & 53856.5491 & 16.67 & -- & 1 & BXS & C \\
53857.2975 & 53857.3700 & 16.246 & 0.011 & 64 & CrAO & V \\
53865.3793 & 53865.3962 & 16.340 & 0.063 & 6 & CrAO & R \\
53866.3781 & 53866.3893 & 16.336 & 0.060 & 6 & CrAO & R \\
53869.4136 & 53869.4342 & 16.481 & 0.032 & 10 & CrAO & R \\
\hline
\end{tabular}
\end{center}
\end{table*}

\section{Results and Discussions}

\subsection{Rebrightenings}\label{sec:reb}

\begin{figure*}
  \begin{center}
    \FigureFile(170mm,80mm){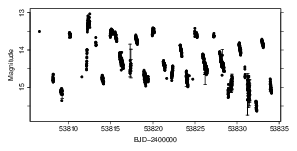}
    \FigureFile(170mm,140mm){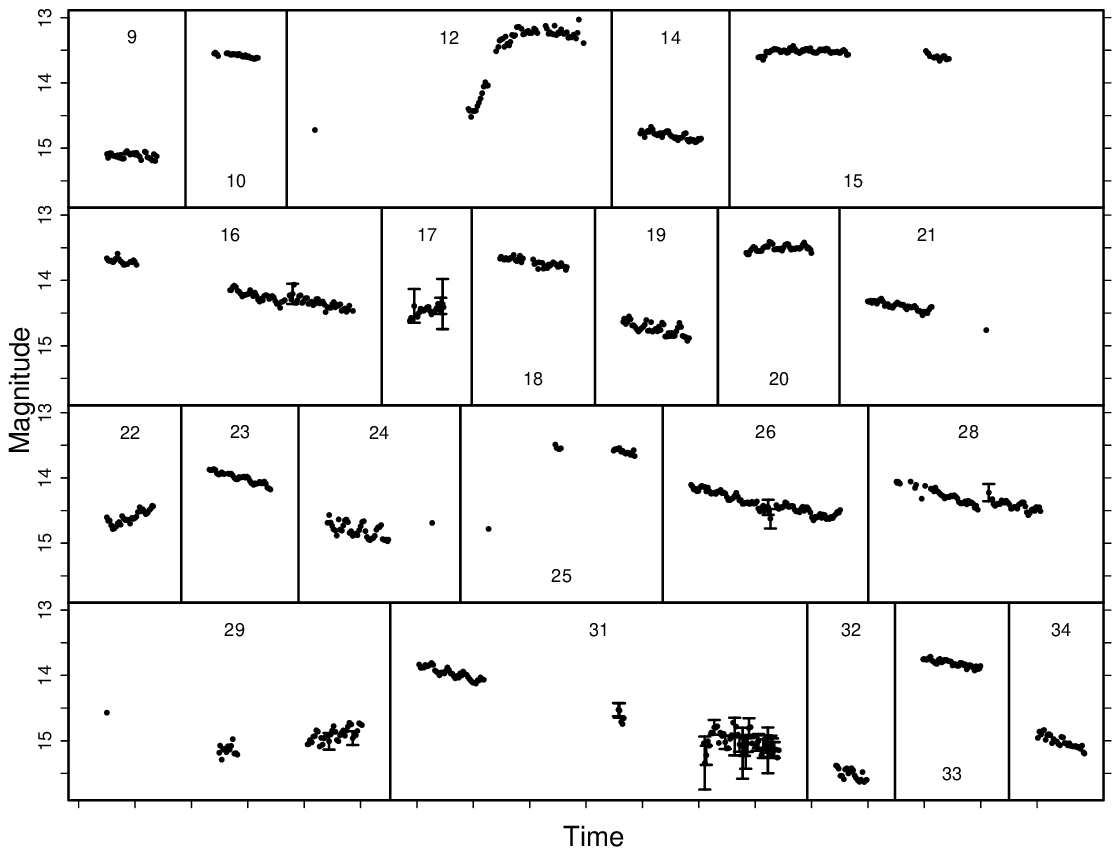}
  \end{center}
  \caption{Light curve of SDSS J0804 during the rebrightening phase.
  (Upper:) Overall light curve.  Eleven rebrightenings are clearly
  visible.
  (Lower:) Enlarged light curves showing hump features.  The numbers
  in each panels indicate truncated
  BJD (BJD-2453800).  The intervals of ticks at the lowest panels
  are 0.2 d.}
  \label{fig:j0804reblc}
\end{figure*}

   SDSS J0804 exhibited eleven rebrightenings
(table \ref{tab:j0804reb}).  During the rebrightenings,
hump features (a combination of superhumps and orbital humps, see
subsection \ref{sec:rebhump} for details) were clearly observed
(figure \ref{fig:j0804reblc}).  As discussed in \citet{pav09j0804},
these rebrightenings showed faster rise and slower decline,
suggesting that they are outside-in type, dwarf nova-type outbursts.
The overall feature of the rebrightenings was similar to those
of EG Cnc although the mean intervals of rebrightenings was
much shorter in SDSS J0804 (2.6(4) d) than in EG Cnc (6.9(3) d).

\begin{table}
\caption{Rebrightenings of SDSS J0804.}\label{tab:j0804reb}
\begin{center}
\begin{tabular}{cc}
\hline\hline
BJD$-$2400000 & Magnitude \\
\hline
53806.5\commenta & 13.9 \\
53810.1 & 13.4 \\
53812.4 & 13.5 \\
53815.2 & 13.4 \\
53817.9 & 13.6 \\
53820.0 & 13.4 \\
53823.2\commentb & 14.1 \\
53825.2 & 13.5 \\
53827.2\commentb & 13.9 \\
53830.2\commentb & 14.1 \\
53832.9 & 13.7 \\
\hline
  \multicolumn{2}{l}{\commenta Single observation.} \\
  \multicolumn{2}{l}{\commentb Maximum not observed.} \\
\end{tabular}
\end{center}
\end{table}

\subsection{Orbital Period}

\begin{figure}
  \begin{center}
    \FigureFile(88mm,110mm){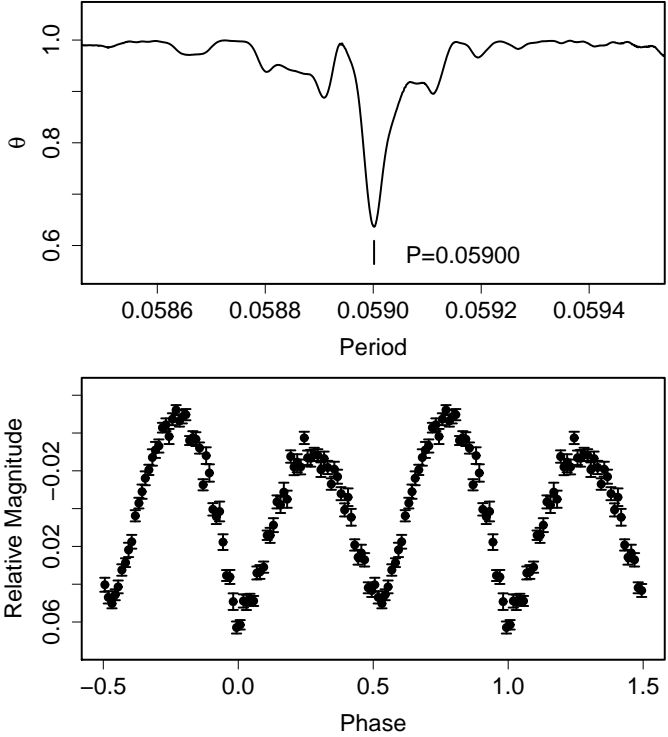}
  \end{center}
  \caption{Period analysis of SDSS J0804 in quiescence.
     (Upper): PDM analysis.
     (Lower): Phase-average profile.
     The phase zero refers to equation \ref{equ:j0804ecl}.}
  \label{fig:j0804quipdm}
\end{figure}

   We analyzed the quiescent data in \citet{zha08j0804}.
A Phase Dispersion Minimization (PDM, \cite{PDM}) period analysis of
the de-trended data clearly indicates the presence
of a single very stable period of 0.0590016(4) d
(figure \ref{fig:j0804quipdm}).  We basically confirmed the results in
\citet{zha08j0804} and identified this period to be the orbital
period ($P_{\rm orb}$).

\begin{figure}
  \begin{center}
    \FigureFile(88mm,110mm){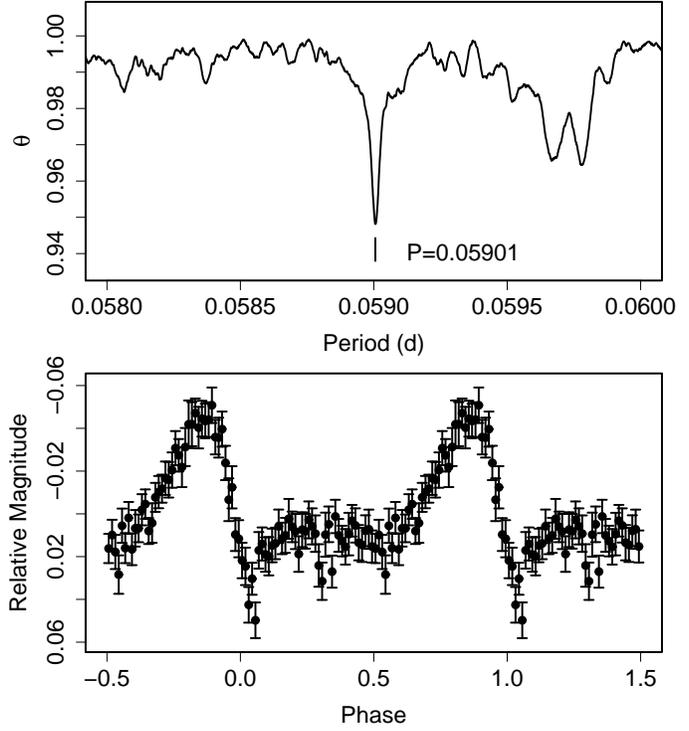}
  \end{center}
  \caption{Period analysis of SDSS J0804 during rebrightenings.
     (Upper): PDM analysis.
     (Lower): Phase-average profile.
     The phase zero refers to equation \ref{equ:j0804ecl}.}
  \label{fig:j0804rebpdm}
\end{figure}

   We then analyzed the stage of rebrightenings after removing the global
trend of rebrightenings.  A period analysis yielded a stable period
of 0.059006(2) d and a broader signal arising from superhumps
with periods around 0.0597--0.0598 d (figure \ref{fig:j0804rebpdm}).
The phase-averaged profile shows a strong hump and a shallow dip-like
fading following the hump.  By analogy with WZ Sge \citep{pat02wzsge},
we identified this dip-like fading as shallow eclipses.  We measured
the epoch of photometric minimum of BJD 2453820.3705(5) based on the
phase-averaged light curve.  The system does not appear to show
a distinct sharp eclipse of the hot spot as in WZ Sge
(cf. \cite{pat02wzsge}).

   By using epochs of eclipses during rebrightenings and final fading,
we identified one of two minima with a mean epoch BJD 2454114.2150(6)
in quiescence as likely eclipses based on analogy with quiescent
light curves of WZ Sge \citep{pat98wzsge} and V455 And
\citep{ara05v455and}.  The times of eclipses (including those recored
during superoutburst, see subsection \ref{sec:eclsuper}) are listed in
table \ref{tab:j0804ecl}.  There possibly
remains cycle 0.5 or 1 ambiguity in selecting the eclipse
in quiescence.

   A direct PDM analysis of the combined data during the rebrightening,
post-superoutburst, and quiescent phases yielded a mean period
of 0.0590031(5) d.

\begin{table}
\caption{Eclipse Minima of SDSS J0804.}\label{tab:j0804ecl}
\begin{center}
\begin{tabular}{cccc}
\hline\hline
$E$ & Minimum\commenta & error & $O-C$\commentb \\
\hline
0 & 53799.3607 & 0.0005 & $-$0.0047 \\
1 & 53799.4200 & 0.0005 & $-$0.0044 \\
2 & 53799.4786 & 0.0008 & $-$0.0048 \\
356 & 53820.3705 & 0.0005 & $-$0.0006 \\
924 & 53853.8865 & 0.0005 & 0.0007 \\
5336 & 54114.2150 & 0.0005 & $-$0.0010 \\
\hline
  \multicolumn{4}{l}{\commenta BJD$-$2400000.} \\
  \multicolumn{4}{l}{\commentb Against equation \ref{equ:j0804ecl}.} \\
\end{tabular}
\end{center}
\end{table}

\subsection{Eclipses during Superoutburst Plateau}\label{sec:eclsuper}

\begin{figure*}
  \begin{center}
    \FigureFile(130mm,60mm){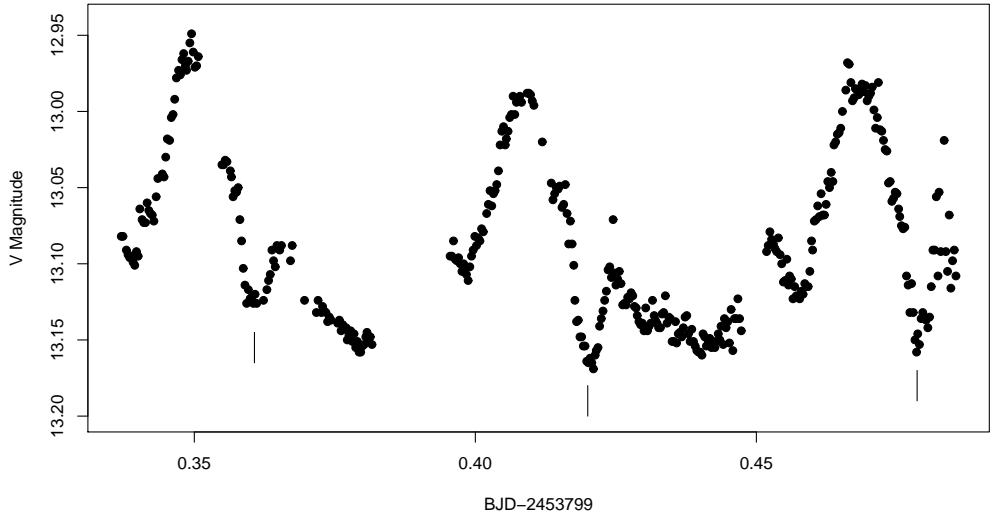}
  \end{center}
  \caption{Light curve of SDSS J0804 on 2006 March 4--5, late stage
  of the superoutburst plateau.  Recurring eclipses (ticks) are present.}
  \label{fig:j0804lcday1}
\end{figure*}

   The light curve on the first night of observation (2006 March 4--5,
late stage of the superoutburst plateau, figure \ref{fig:j0804lcday1})
clearly shows the presence of recurring eclipses other than superhumps.
These eclipses in outburst were prominently seen on 2006 March 4--5,
but were almost absent on March 5--7 (during the stage of
rapid fading).

   There was a systematic 0.004--0.005 d difference
between epochs of eclipses during superoutburst plateau and rebrightenings
(cf. table \ref{tab:j0804ecl}).
This difference can be understood if a different portion of the accretion
disk eclipsed during different states.  We disregarded eclipses
during superoutbursts in determining the orbital ephemeris
(equation \ref{equ:j0804ecl}), since this period better represents
light curves during the most of the observed epochs than the period
determined from all eclipses.
The given ephemeris thus has this degree of uncertainty, and needs to be
verified by further observations.
This uncertainty does not affect the following analysis of superhumps.

\begin{equation}
{\rm Min(BJD)} = 2453799.3654(7) + 0.0590048(2)
\label{equ:j0804ecl}.
\end{equation}

   The appearance of distinct eclipses only during the superoutburst
plateau can be understood as a result of the luminous portion (hot state)
of the accretion disk expanding during the outburst as predicted by
the disk-instability theory.  The system is thus a grazing eclipser
whose eclipses are only prominent when the radius of the accretion disk
sufficiently expands.

\subsection{Superhumps during the Main Superoutburst}

   We measured times of superhump maxima outside the eclipses after
removing observations within 0.07 $P_{\rm orb}$ of eclipses.
The analyzed data covered the last part of
the superoutburst plateau and the rapid fading stage
(table \ref{tab:j0804oc2006}).  It is evident that the times of
superhump maxima can not be expressed by a single constant period.
The difference in $O-C$ between $E=40$ and $E=44$ is too large to be
considered as a real period change.  Since all the maxima for
$44 \le E \le 49$ have orbital phases of 0.70--0.79, they most
likely represent a projection effect in a high-inclination system
(\cite{osa03DNoutburst} figures 2 and 3; Kato et al., in preparation;
we call them orbital humps in this paper for simplicity).

   We disregarded this portion and obtained the mean superhump period
($P_{\rm SH}$) of 0.05954(3) d ($E \le 40$)
from the times of maxima.  A PDM analysis of the corresponding
segment of the data yielded a period of 0.059539(11) d, and we
adopt this as being the representative $P_{\rm SH}$ of this object.
The fractional superhump excess $\epsilon = P_{\rm SH}/P_{\rm orb}-1$
for this period is 0.90(2) \%.  As in most of WZ Sge-type dwarf novae
(Kato et al., in preparation), a shortening of the superhump period
toward the end of the superoutburst as frequently seen in the
majority of ordinary SU UMa-type dwarf novae, was absent.
The present period is significantly shorter than
the $P_{\rm SH}$ of 0.059713(7) d \citet{pav07j0804}.
This difference was caused by the incorrect times of observations
for the first night used in \citet{pav07j0804}.
The period agrees 0.0597(11) d by \citet{she07j0804} within their
error.

\begin{figure}
  \begin{center}
    \FigureFile(88mm,110mm){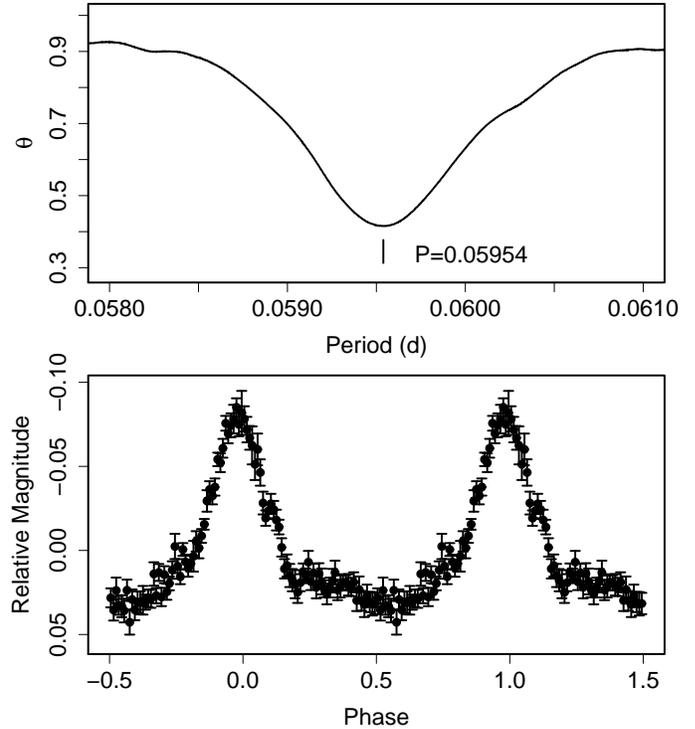}
  \end{center}
  \caption{Period analysis of SDSS J0804 during the main superoutburst.
     (Upper): PDM analysis.
     (Lower): Phase-average profile.
     The phase zero refers to equation \ref{equ:j0804ecl}.}
  \label{fig:j0804mainpdm}
\end{figure}

\begin{table}
\caption{Superhump Maxima of SDSS J0804 (2006).}\label{tab:j0804oc2006}
\begin{center}
\begin{tabular}{cccccc}
\hline\hline
$E$ & max$^a$ & error & $O-C^b$ & phase$^c$ & $N^d$ \\
\hline
0 & 53799.3495 & 0.0002 & $-$0.0050 & 0.73 & 64 \\
1 & 53799.4081 & 0.0002 & $-$0.0056 & 0.72 & 71 \\
2 & 53799.4679 & 0.0005 & $-$0.0049 & 0.74 & 93 \\
17 & 53800.3611 & 0.0003 & 0.0009 & 0.88 & 120 \\
18 & 53800.4204 & 0.0004 & 0.0010 & 0.88 & 93 \\
19 & 53800.4794 & 0.0005 & 0.0009 & 0.88 & 88 \\
20 & 53800.5380 & 0.0004 & 0.0003 & 0.87 & 89 \\
21 & 53800.5997 & 0.0005 & 0.0029 & 0.92 & 66 \\
22 & 53800.6587 & 0.0006 & 0.0027 & 0.92 & 29 \\
23 & 53800.7142 & 0.0018 & $-$0.0009 & 0.86 & 11 \\
34 & 53801.3769 & 0.0009 & 0.0110 & 0.09 & 22 \\
35 & 53801.4313 & 0.0011 & 0.0062 & 0.01 & 43 \\
37 & 53801.5517 & 0.0011 & 0.0083 & 0.05 & 67 \\
38 & 53801.6114 & 0.0041 & 0.0088 & 0.06 & 82 \\
39 & 53801.6706 & 0.0016 & 0.0089 & 0.07 & 14 \\
40 & 53801.7292 & 0.0010 & 0.0083 & 0.06 & 14 \\
44 & 53801.9438 & 0.0020 & $-$0.0137 & 0.70 & 65 \\
47 & 53802.1262 & 0.0013 & $-$0.0088 & 0.79 & 61 \\
48 & 53802.1847 & 0.0015 & $-$0.0095 & 0.78 & 74 \\
49 & 53802.2415 & 0.0014 & $-$0.0118 & 0.74 & 74 \\
\hline
  \multicolumn{5}{l}{$^{a}$ BJD$-$2400000.} \\
  \multicolumn{5}{l}{$^{b}$ Against $max = 2453799.3545 + 0.059160 E$.} \\
  \multicolumn{6}{l}{$^{c}$ Orbital phase.} \\
  \multicolumn{6}{l}{$^{d}$ Number of points used to determine the maximum.} \\
\end{tabular}
\end{center}
\end{table}

\subsection{Hump Features during the Rebrightening Phase}\label{sec:rebhump}

\begin{figure}
  \begin{center}
    \FigureFile(88mm,110mm){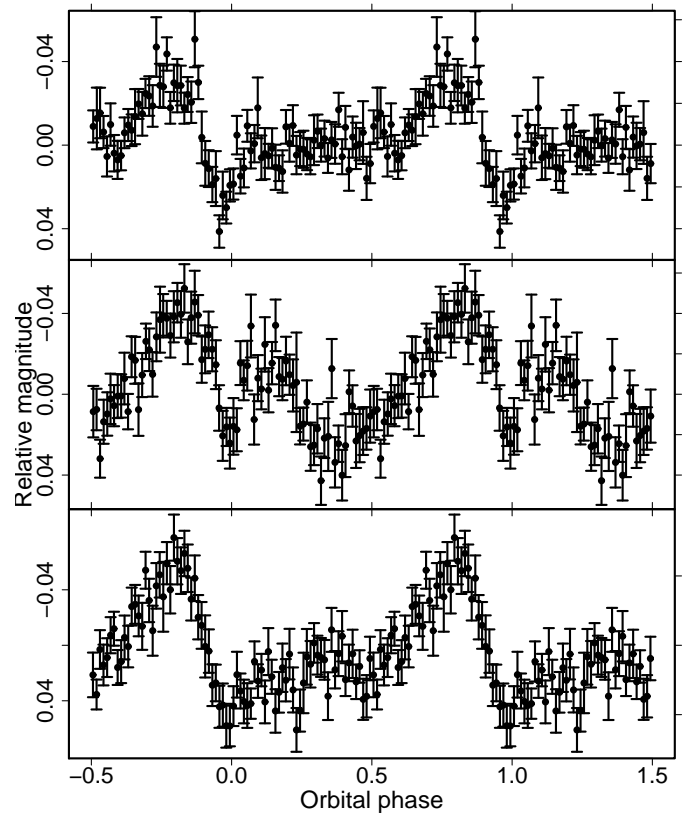}
  \end{center}
  \caption{Phase-averaged light curves during the rebrightening phase
  at different brightness levels.  Upper: near maximum (brighter than
  13.8 mag).  Middle: intermediate brightness (13.8--14.5 mag).
  Lower: near minimum (fainter than 14.5 mag).}
  \label{fig:j0804rebave}
\end{figure}

   Phase-averaged light curves during the rebrightening phase
at different brightness levels are shown in figure \ref{fig:j0804rebave}.
The profiles of orbital variation were similar regardless of the
brightness level.  Since the system luminosity varied by more than
a factor of two between levels, the amplitude of the orbital humps
is expected to vary by a similar factor if the orbital humps (assuming
a constant luminosity) arises from the hot spot (cf. \cite{pat02wzsge})
as in dwarf novae in quiescence.  The relatively constant amplitude of
the orbital humps indicates, on the contrary, that the luminosity of
the orbital humps varies proportionally to the system luminosity.
This result is consistent with a projection effect and is contrary to
what it expected for a hot spot from an enhanced mass-transfer,
strengthening the argument in \citet{osa03DNoutburst}.

\begin{figure}
  \begin{center}
    \FigureFile(88mm,110mm){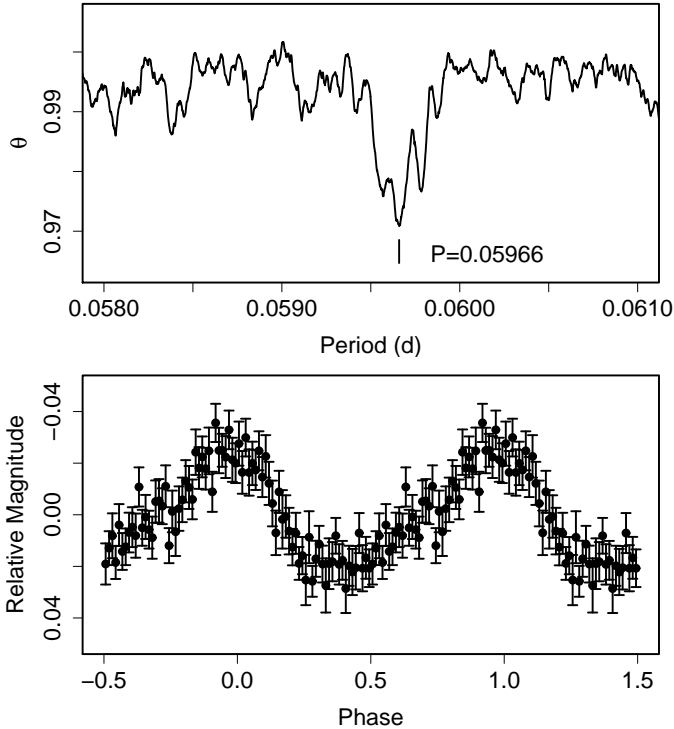}
  \end{center}
  \caption{Period analysis of SDSS J0804 during rebrightenings
     after subtracting the mean orbital variation.
     (Upper): PDM analysis.
     (Lower): Phase-average profile.}
  \label{fig:j0804rebsubpdm}
\end{figure}

   We performed period analysis during the rebrightening stage
after removing the global trend of rebrightenings and subtracting
the mean orbital variation.  A PDM analysis has yielded a strong
superhump signal with a mean period of 0.059659(5) d
(figure \ref{fig:j0804rebsubpdm}).

   We measured times of humps during the rebrightening phase
using two methods.  Since the variation was extremely complex,
we give both results for complementary purposes and for a comparison
with previous works.

   The first method directly used the light curve after
removing the global trend of rebrightenings
(table \ref{tab:j0804ochump}).  These maxima correspond to the times
presented in \citet{pav09j0804}.
Since these times consisted of a mixture of orbital humps and superhumps,
we selected maxima for their orbital phases
$0 < \rm{phase} < 0.6$.  The selected maxima for $E \ge 219$ can be
very well expressed by a single constant period of 0.059631(11) d.

   The second method used the light curve subtracted for the mean
orbital light curve during the rebrightening.  This method is expected
to be more sensitive to superhumps, while there remains a possibility
of the effect of the variation in the orbital light curve.
The extracted times are listed in table
\ref{tab:j0804ochump2}.  For the interval $E \le 135$ and
$236 \le E \le 536$, the $P_{\rm SH}$ was almost constant at 0.059632(6) d.
We regard this period as the representative $P_{\rm SH}$ during the
rebrightening phase.
The period is 0.15(3) \% longer than the $P_{\rm SH}$ during
the superoutburst plateau.  The unusual cyclic $O-C$ variation
reported in \citet{pav09j0804} was probably caused by a contamination
by orbital humps.

\begin{table}
\caption{Maxima of humps during rebrightening phase of SDSS J0804.}\label{tab:j0804ochump}
\begin{center}
\begin{tabular}{cccccc}
\hline\hline
$E$ & max$^a$ & error & $O-C^b$ & phase$^c$ & $N^d$ \\
\hline
0 & 53808.1456 & 0.0012 & $-$0.0280 & 0.80 & 92 \\
16 & 53809.1186 & 0.0017 & $-$0.0109 & 0.29 & 93 \\
67 & 53812.1589 & 0.0006 & $-$0.0175 & 0.82 & 82 \\
97 & 53813.9750 & 0.0016 & 0.0063 & 0.60 & 81 \\
98 & 53814.0333 & 0.0010 & 0.0049 & 0.59 & 94 \\
99 & 53814.0945 & 0.0010 & 0.0063 & 0.62 & 95 \\
113 & 53814.9276 & 0.0011 & 0.0031 & 0.75 & 43 \\
114 & 53814.9797 & 0.0026 & $-$0.0046 & 0.63 & 66 \\
115 & 53815.0426 & 0.0013 & $-$0.0014 & 0.69 & 66 \\
116 & 53815.0963 & 0.0026 & $-$0.0074 & 0.60 & 66 \\
117 & 53815.1632 & 0.0013 & $-$0.0003 & 0.74 & 145 \\
118 & 53815.2211 & 0.0011 & $-$0.0021 & 0.72 & 119 \\
124 & 53815.5799 & 0.0005 & $-$0.0018 & 0.80 & 7 \\
125 & 53815.6401 & 0.0016 & $-$0.0013 & 0.82 & 7 \\
126 & 53815.6949 & 0.0040 & $-$0.0062 & 0.75 & 7 \\
132 & 53816.0552 & 0.0011 & $-$0.0044 & 0.86 & 91 \\
133 & 53816.1142 & 0.0023 & $-$0.0052 & 0.85 & 98 \\
134 & 53816.1727 & 0.0021 & $-$0.0064 & 0.85 & 114 \\
135 & 53816.2305 & 0.0025 & $-$0.0083 & 0.83 & 53 \\
136 & 53816.2897 & 0.0022 & $-$0.0089 & 0.83 & 16 \\
137 & 53816.3504 & 0.0016 & $-$0.0079 & 0.86 & 16 \\
138 & 53816.4084 & 0.0028 & $-$0.0097 & 0.84 & 16 \\
164 & 53817.9919 & 0.0027 & 0.0205 & 0.68 & 33 \\
166 & 53818.1102 & 0.0034 & 0.0194 & 0.68 & 59 \\
180 & 53818.9415 & 0.0013 & 0.0142 & 0.77 & 64 \\
181 & 53818.9984 & 0.0011 & 0.0114 & 0.74 & 65 \\
182 & 53819.0564 & 0.0013 & 0.0097 & 0.72 & 66 \\
183 & 53819.1184 & 0.0007 & 0.0120 & 0.77 & 60 \\
197 & 53819.9420 & 0.0007 & $-$0.0009 & 0.73 & 64 \\
198 & 53820.0015 & 0.0011 & $-$0.0011 & 0.74 & 66 \\
199 & 53820.0620 & 0.0007 & $-$0.0003 & 0.76 & 56 \\
200 & 53820.1226 & 0.0008 & 0.0005 & 0.79 & 58 \\
219 & 53821.2671 & 0.0007 & 0.0099 & 0.18 & 16 \\
220 & 53821.3327 & 0.0025 & 0.0158 & 0.30 & 17 \\
221 & 53821.3826 & 0.0027 & 0.0059 & 0.14 & 17 \\
236 & 53822.2765 & 0.0030 & 0.0036 & 0.29 & 17 \\
237 & 53822.3364 & 0.0010 & 0.0038 & 0.31 & 16 \\
252 & 53823.2432 & 0.0011 & 0.0145 & 0.68 & 15 \\
253 & 53823.3027 & 0.0009 & 0.0143 & 0.68 & 13 \\
254 & 53823.3565 & 0.0018 & 0.0084 & 0.60 & 17 \\
255 & 53823.4205 & 0.0024 & 0.0126 & 0.68 & 17 \\
264 & 53823.9561 & 0.0012 & 0.0105 & 0.76 & 38 \\
265 & 53824.0150 & 0.0004 & 0.0097 & 0.76 & 154 \\
266 & 53824.0739 & 0.0004 & 0.0088 & 0.75 & 156 \\
267 & 53824.1320 & 0.0006 & 0.0071 & 0.74 & 151 \\
285 & 53825.2000 & 0.0019 & $-$0.0002 & 0.84 & 90 \\
298 & 53825.9695 & 0.0012 & $-$0.0073 & 0.88 & 65 \\
299 & 53826.0334 & 0.0030 & $-$0.0032 & 0.96 & 136 \\
300 & 53826.0867 & 0.0012 & $-$0.0097 & 0.87 & 125 \\
301 & 53826.1500 & 0.0034 & $-$0.0061 & 0.94 & 59 \\
303 & 53826.2659 & 0.0016 & $-$0.0096 & 0.90 & 17 \\
304 & 53826.3353 & 0.0015 & $-$0.0000 & 0.08 & 17 \\
\hline
  \multicolumn{6}{l}{$^{a}$ BJD$-$2400000.} \\
  \multicolumn{6}{l}{$^{b}$ Against $max = 2453808.1736 + 0.059742 E$.} \\
  \multicolumn{6}{l}{$^{c}$ Orbital phase.} \\
  \multicolumn{6}{l}{$^{d}$ Number of points used to determine the maximum.} \\
\end{tabular}
\end{center}
\end{table}

\addtocounter{table}{-1}
\begin{table}
\caption{Maxima of humps during rebrightening phase of SDSS J0804 (continued).}
\begin{center}
\begin{tabular}{cccccc}
\hline\hline
$E$ & max & error & $O-C$ & phase & $N$ \\
\hline
306 & 53826.4543 & 0.0044 & $-$0.0005 & 0.10 & 11 \\
334 & 53828.1287 & 0.0016 & 0.0012 & 0.47 & 89 \\
335 & 53828.1837 & 0.0037 & $-$0.0036 & 0.41 & 95 \\
337 & 53828.3096 & 0.0047 & 0.0028 & 0.54 & 16 \\
338 & 53828.3699 & 0.0012 & 0.0033 & 0.56 & 17 \\
339 & 53828.4229 & 0.0042 & $-$0.0034 & 0.46 & 15 \\
348 & 53828.9679 & 0.0020 & 0.0040 & 0.70 & 65 \\
353 & 53829.2660 & 0.0014 & 0.0033 & 0.75 & 16 \\
354 & 53829.3285 & 0.0008 & 0.0061 & 0.81 & 16 \\
355 & 53829.3840 & 0.0018 & 0.0019 & 0.75 & 17 \\
370 & 53830.2707 & 0.0012 & $-$0.0077 & 0.78 & 16 \\
371 & 53830.3314 & 0.0009 & $-$0.0066 & 0.81 & 17 \\
372 & 53830.3882 & 0.0017 & $-$0.0096 & 0.77 & 17 \\
373 & 53830.4485 & 0.0021 & $-$0.0091 & 0.79 & 14 \\
387 & 53831.2786 & 0.0025 & $-$0.0154 & 0.86 & 17 \\
388 & 53831.3476 & 0.0028 & $-$0.0061 & 0.03 & 17 \\
389 & 53831.4050 & 0.0011 & $-$0.0084 & 0.00 & 15 \\
390 & 53831.4615 & 0.0030 & $-$0.0116 & 0.96 & 16 \\
403 & 53832.2865 & 0.0018 & 0.0367 & 0.94 & 46 \\
404 & 53832.3412 & 0.0013 & 0.0317 & 0.87 & 42 \\
415 & 53832.9544 & 0.0013 & $-$0.0123 & 0.26 & 63 \\
416 & 53833.0169 & 0.0022 & $-$0.0096 & 0.32 & 66 \\
417 & 53833.0708 & 0.0016 & $-$0.0154 & 0.23 & 66 \\
433 & 53834.0256 & 0.0022 & $-$0.0164 & 0.41 & 121 \\
434 & 53834.0844 & 0.0085 & $-$0.0175 & 0.41 & 115 \\
\hline
\end{tabular}
\end{center}
\end{table}

\begin{table}
\caption{Maxima of humps during rebrightening phase of SDSS J0804 after subtracting the orbital signal.}\label{tab:j0804ochump2}
\begin{center}
\begin{tabular}{cccccc}
\hline\hline
$E$ & max$^a$ & error & $O-C^b$ & phase$^c$ & $N^d$ \\
\hline
0 & 53802.1248 & 0.0017 & 0.0116 & 0.77 & 75 \\
1 & 53802.1844 & 0.0025 & 0.0117 & 0.78 & 95 \\
2 & 53802.2373 & 0.0019 & 0.0049 & 0.67 & 89 \\
101 & 53808.1497 & 0.0044 & 0.0125 & 0.87 & 92 \\
117 & 53809.1183 & 0.0017 & 0.0268 & 0.29 & 93 \\
119 & 53809.2167 & 0.0022 & 0.0059 & 0.96 & 35 \\
134 & 53810.1094 & 0.0033 & 0.0040 & 0.09 & 82 \\
135 & 53810.1744 & 0.0035 & 0.0092 & 0.19 & 136 \\
199 & 53813.9717 & 0.0010 & $-$0.0106 & 0.54 & 80 \\
200 & 53814.0305 & 0.0007 & $-$0.0115 & 0.54 & 94 \\
201 & 53814.0915 & 0.0009 & $-$0.0101 & 0.58 & 94 \\
215 & 53814.9251 & 0.0018 & $-$0.0116 & 0.70 & 36 \\
216 & 53814.9742 & 0.0012 & $-$0.0221 & 0.53 & 65 \\
217 & 53815.0368 & 0.0026 & $-$0.0192 & 0.59 & 65 \\
218 & 53815.0894 & 0.0016 & $-$0.0262 & 0.49 & 66 \\
219 & 53815.1454 & 0.0046 & $-$0.0298 & 0.44 & 113 \\
220 & 53815.2016 & 0.0050 & $-$0.0333 & 0.39 & 115 \\
234 & 53816.0594 & 0.0016 & $-$0.0105 & 0.93 & 91 \\
235 & 53816.1210 & 0.0015 & $-$0.0085 & 0.97 & 111 \\
236 & 53816.1886 & 0.0016 & $-$0.0006 & 0.12 & 107 \\
237 & 53816.2631 & 0.0029 & 0.0143 & 0.38 & 15 \\
238 & 53816.3061 & 0.0030 & $-$0.0024 & 0.11 & 16 \\
239 & 53816.3579 & 0.0027 & $-$0.0102 & 0.99 & 16 \\
240 & 53816.4215 & 0.0019 & $-$0.0063 & 0.06 & 15 \\
255 & 53817.3310 & 0.0016 & 0.0085 & 0.48 & 17 \\
256 & 53817.3738 & 0.0122 & $-$0.0083 & 0.20 & 15 \\
268 & 53818.1049 & 0.0027 & 0.0071 & 0.59 & 54 \\
283 & 53818.9976 & 0.0016 & 0.0052 & 0.72 & 66 \\
284 & 53819.0559 & 0.0014 & 0.0037 & 0.71 & 65 \\
285 & 53819.1185 & 0.0012 & 0.0068 & 0.77 & 61 \\
299 & 53819.9398 & 0.0015 & $-$0.0070 & 0.69 & 65 \\
300 & 53819.9984 & 0.0035 & $-$0.0080 & 0.68 & 66 \\
301 & 53820.0626 & 0.0017 & $-$0.0034 & 0.77 & 56 \\
302 & 53820.1245 & 0.0014 & $-$0.0012 & 0.82 & 54 \\
322 & 53821.3304 & 0.0016 & 0.0118 & 0.26 & 17 \\
323 & 53821.3830 & 0.0015 & 0.0047 & 0.15 & 17 \\
324 & 53821.4504 & 0.0057 & 0.0125 & 0.29 & 9 \\
338 & 53822.2762 & 0.0037 & 0.0032 & 0.29 & 16 \\
339 & 53822.3363 & 0.0011 & 0.0037 & 0.31 & 16 \\
355 & 53823.2853 & 0.0071 & $-$0.0016 & 0.39 & 12 \\
356 & 53823.3510 & 0.0028 & 0.0045 & 0.50 & 16 \\
357 & 53823.4131 & 0.0026 & 0.0070 & 0.56 & 16 \\
369 & 53824.1318 & 0.0007 & 0.0099 & 0.74 & 150 \\
400 & 53825.9798 & 0.0036 & 0.0089 & 0.05 & 65 \\
402 & 53826.0913 & 0.0017 & 0.0011 & 0.94 & 114 \\
403 & 53826.1551 & 0.0028 & 0.0053 & 0.03 & 60 \\
405 & 53826.2729 & 0.0022 & 0.0038 & 0.02 & 17 \\
406 & 53826.3354 & 0.0017 & 0.0066 & 0.08 & 17 \\
\hline
  \multicolumn{6}{l}{$^{a}$ BJD$-$2400000.} \\
  \multicolumn{6}{l}{$^{b}$ Against $max = 2453802.1131 + 0.059644 E$.} \\
  \multicolumn{6}{l}{$^{c}$ Orbital phase.} \\
  \multicolumn{6}{l}{$^{d}$ Number of points used to determine the maximum.} \\
\end{tabular}
\end{center}
\end{table}

\addtocounter{table}{-1}
\begin{table}
\caption{Maxima of humps during rebrightening phase of SDSS J0804 after subtracting the orbital signal (continued).}
\begin{center}
\begin{tabular}{cccccc}
\hline\hline
$E$ & max & error & $O-C$ & phase & $N$ \\
\hline
408 & 53826.4518 & 0.0022 & 0.0038 & 0.06 & 9 \\
421 & 53827.2227 & 0.0018 & $-$0.0007 & 0.12 & 40 \\
436 & 53828.1258 & 0.0011 & 0.0078 & 0.43 & 90 \\
437 & 53828.1822 & 0.0023 & 0.0045 & 0.38 & 94 \\
439 & 53828.3047 & 0.0016 & 0.0077 & 0.46 & 17 \\
440 & 53828.3677 & 0.0010 & 0.0110 & 0.52 & 17 \\
441 & 53828.4198 & 0.0035 & 0.0035 & 0.41 & 17 \\
450 & 53828.9581 & 0.0064 & 0.0050 & 0.53 & 63 \\
456 & 53829.3072 & 0.0027 & $-$0.0037 & 0.45 & 17 \\
457 & 53829.3677 & 0.0033 & $-$0.0029 & 0.47 & 17 \\
458 & 53829.4244 & 0.0008 & $-$0.0059 & 0.43 & 12 \\
472 & 53830.2630 & 0.0079 & $-$0.0022 & 0.65 & 17 \\
474 & 53830.3871 & 0.0050 & 0.0026 & 0.75 & 17 \\
475 & 53830.4470 & 0.0039 & 0.0028 & 0.76 & 17 \\
489 & 53831.2823 & 0.0026 & 0.0031 & 0.92 & 17 \\
490 & 53831.3489 & 0.0024 & 0.0101 & 0.05 & 16 \\
491 & 53831.4056 & 0.0011 & 0.0071 & 0.01 & 15 \\
492 & 53831.4646 & 0.0049 & 0.0065 & 0.01 & 17 \\
506 & 53832.2947 & 0.0019 & 0.0016 & 0.08 & 47 \\
517 & 53832.9537 & 0.0018 & 0.0045 & 0.25 & 63 \\
518 & 53833.0164 & 0.0022 & 0.0075 & 0.31 & 65 \\
519 & 53833.0700 & 0.0020 & 0.0015 & 0.22 & 65 \\
534 & 53833.9542 & 0.0021 & $-$0.0090 & 0.20 & 104 \\
535 & 53834.0253 & 0.0016 & 0.0025 & 0.41 & 120 \\
536 & 53834.0829 & 0.0054 & 0.0004 & 0.38 & 114 \\
573 & 53836.2836 & 0.0074 & $-$0.0058 & 0.68 & 8 \\
623 & 53839.2464 & 0.0038 & $-$0.0251 & 0.89 & 17 \\
625 & 53839.3807 & 0.0026 & $-$0.0102 & 0.17 & 14 \\
\hline
\end{tabular}
\end{center}
\end{table}

\subsection{Final Fading Stage}

   After a sequence of eleven rebrightenings, this object entered
the stage of final slow fading.  In contrast to EG Cnc
(\cite{pat98egcnc}; \cite{kat04egcnc}), the fading trend of the
final fading was on a smooth extension of the minima during the
rebrightening phase.  \citet{osa97egcnc}, \citet{osa01egcnc} interpreted
the jump in EG Cnc as representing the decrease in the viscosity
in the accretion disk.  The apparent lack of this feature in
SDSS J0804 suggests that a stepwise decrease in the viscosity is not
required to terminate the rebrightening activity.

\begin{figure}
  \begin{center}
    \FigureFile(88mm,70mm){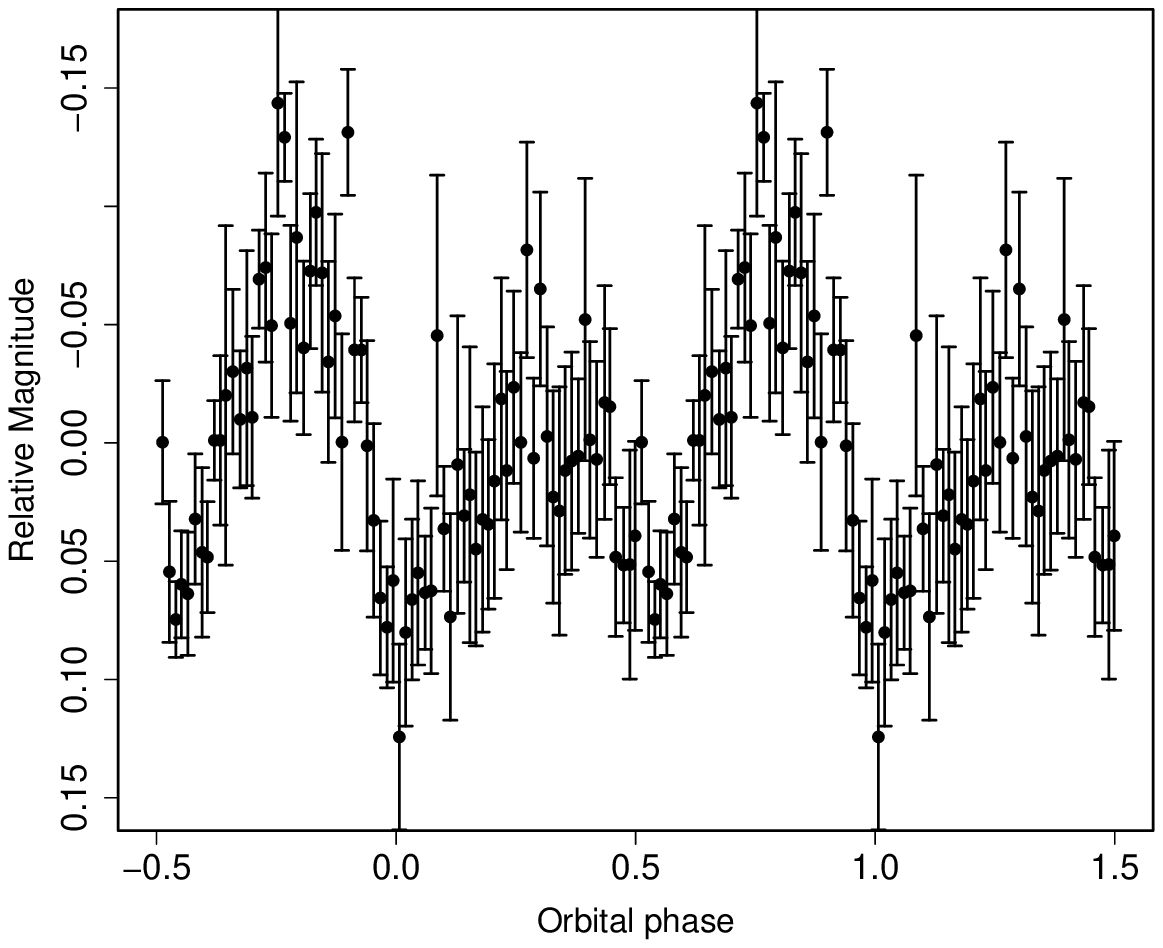}
  \end{center}
  \caption{Phase-averaged orbital Light curve of SDSS J0804 during
  2006 April 21--May 13 (post-superoutburst final fading).}
  \label{fig:j0804postave}
\end{figure}

\begin{figure}
  \begin{center}
    \FigureFile(88mm,110mm){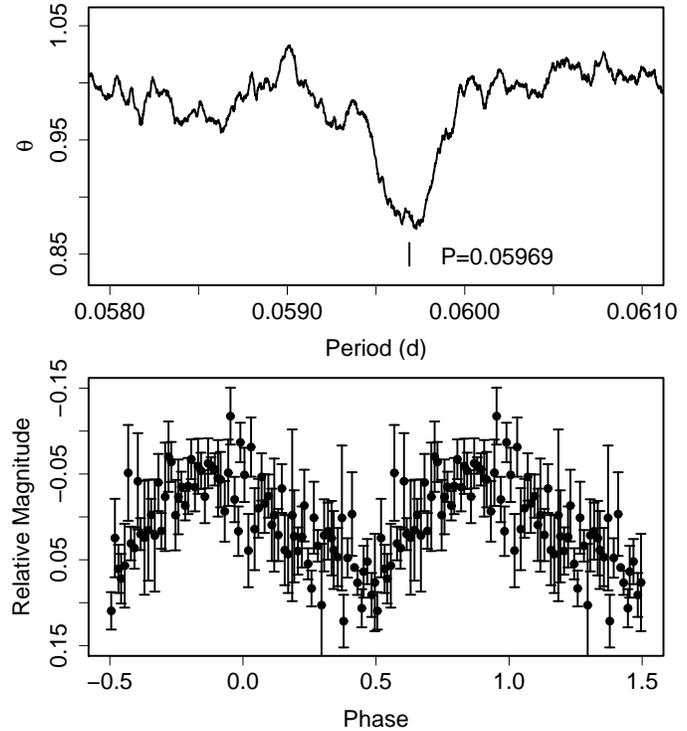}
  \end{center}
  \caption{Period analysis of SDSS J0804 during the final fading phase
     after subtracting the mean orbital variation.
     (Upper): PDM analysis.
     (Lower): Phase-average profile.}
  \label{fig:j0804postsubpdm}
\end{figure}

   A phase-averaged orbital light curve of SDSS J0804 during
2006 April 21--May 13 is shown in in figure \ref{fig:j0804postave}.
Although the orbital hump remained strong, the eclipse became
less prominent.  The reduction of the eclipse feature suggests
that the luminous portion of the accretion disk had shrunk compared
to the rebrightening phase.  After subtracting the mean orbital
light curve, the superhump signal with a period of 0.05969(4) d,
0.25(6) \% longer than the superhump period during the superoutburst
plateau, was strongly detected
(figure \ref{fig:j0804postsubpdm}; table \ref{tab:j0804ocpost}).
This period suggests that the superhump period further lengthened
during the post-rebrightening phase.

\begin{figure*}
  \begin{center}
    \FigureFile(160mm,160mm){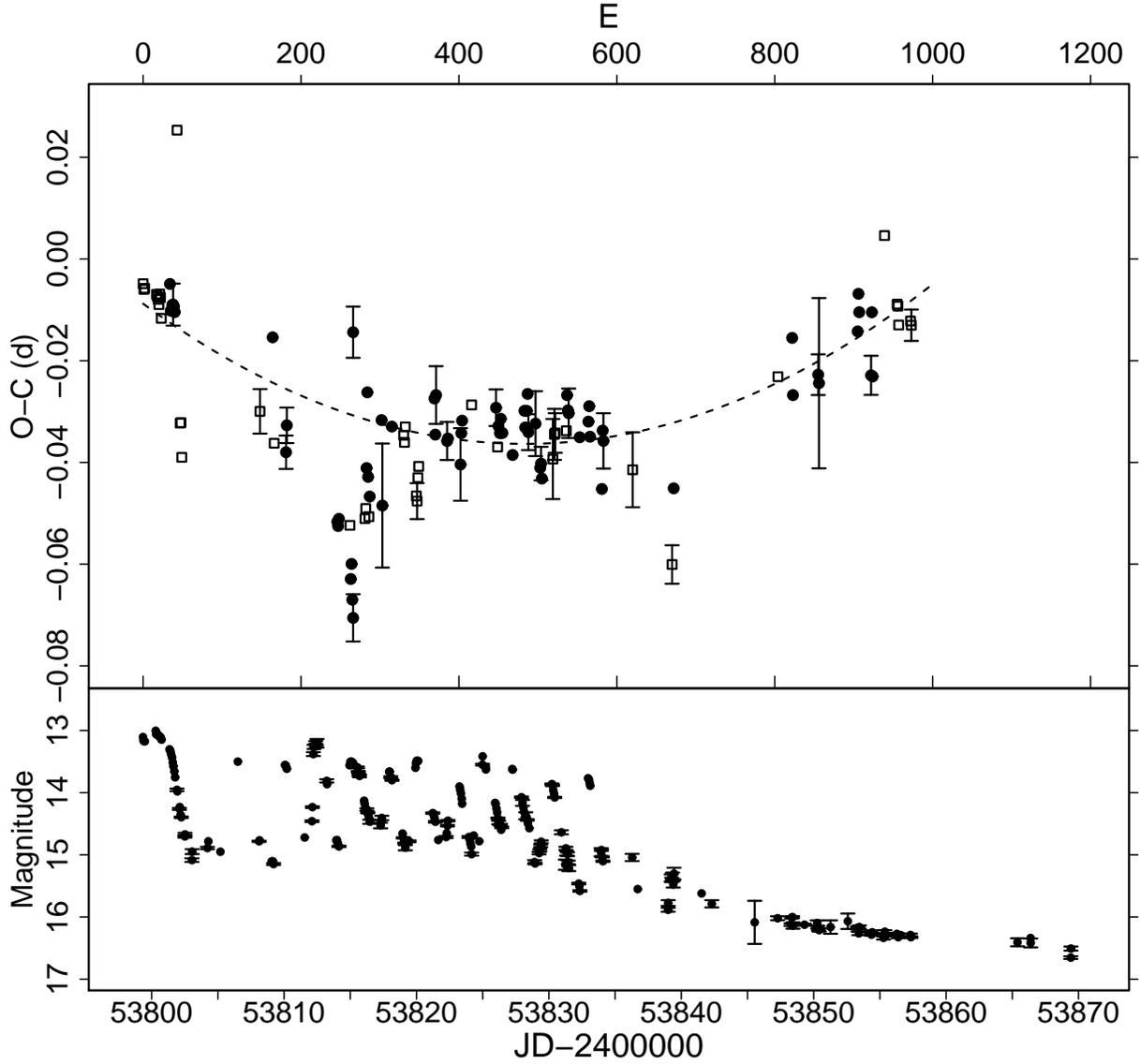}
  \end{center}
  \caption{$O-C$ variation of superhumps in SDSS J0804 during the entire
  course of the outburst.  (Upper) $O-C$.
  Open squares indicate humps coinciding with the phase of orbital humps.
  Filled squares are humps outside the phase of orbital humps.
  We used a period of 0.05963 d for calculating the $O-C$'s.
  During the rebrightening phase, hump times after subtracting the
  orbital variations have been used.
  The period of superhumps continuously increased during the rebrightening
  and post-superoutburst phase.
  The dashed curve represents a quadratic fit with
  $\dot{P}/P$ = $+0.5 \times 10^{-5}$.
  (Lower) Light curve.  Eleven rebrightenings were recorded following
  the superoutburst plateau.  The early stage of the superoutburst
  was not observed.
  }
  \label{fig:j0804humpall}
\end{figure*}

   The $O-C$ variation of superhumps in SDSS J0804 during the entire
course of the outburst is shown in figure \ref{fig:j0804humpall}.
The period of superhumps continuously increased during the rebrightening
and post-superoutburst phase, and the epochs of superhumps during
the superoutburst plateau, rebrightening phase and final fading phase
can be smoothly linked by a period derivative of
$\dot{P}/P$ = $+0.5 \times 10^{-5}$.
The existence of a longer $P_{\rm SH}$ during the post-superoutburst
stage is similar to GW Lib, V455 And and WZ Sge \citep{kat08wzsgelateSH}.

\begin{table}
\caption{Maxima of humps during final fading phase of SDSS J0804}\label{tab:j0804ocpost}
\begin{center}
\begin{tabular}{cccccc}
\hline\hline
$E$ & max$^a$ & error & $O-C^b$ & phase$^c$ & $N^d$ \\
\hline
0 & 53847.2738 & 0.0014 & 0.0006 & 0.94 & 14 \\
18 & 53848.3548 & 0.0025 & 0.0067 & 0.26 & 17 \\
19 & 53848.4031 & 0.0020 & $-$0.0047 & 0.08 & 17 \\
51 & 53850.3153 & 0.0040 & $-$0.0035 & 0.49 & 16 \\
52 & 53850.3733 & 0.0167 & $-$0.0052 & 0.47 & 16 \\
101 & 53853.3053 & 0.0021 & 0.0007 & 0.16 & 16 \\
102 & 53853.3723 & 0.0014 & 0.0080 & 0.30 & 9 \\
103 & 53853.4283 & 0.0015 & 0.0043 & 0.25 & 15 \\
118 & 53854.3104 & 0.0038 & $-$0.0095 & 0.20 & 16 \\
119 & 53854.3824 & 0.0014 & 0.0029 & 0.42 & 17 \\
120 & 53854.4294 & 0.0022 & $-$0.0099 & 0.21 & 8 \\
135 & 53855.3516 & 0.0011 & 0.0165 & 0.84 & 16 \\
151 & 53856.2922 & 0.0006 & 0.0017 & 0.78 & 32 \\
152 & 53856.3514 & 0.0006 & 0.0012 & 0.79 & 27 \\
153 & 53856.4073 & 0.0009 & $-$0.0026 & 0.73 & 18 \\
168 & 53857.3026 & 0.0023 & $-$0.0031 & 0.91 & 21 \\
169 & 53857.3614 & 0.0031 & $-$0.0041 & 0.90 & 23 \\
\hline
  \multicolumn{6}{l}{$^{a}$ BJD$-$2400000.} \\
  \multicolumn{6}{l}{$^{b}$ Against $max = 2453847.2732 + 0.059718 E$.} \\
  \multicolumn{6}{l}{$^{c}$ Orbital phase.} \\
  \multicolumn{6}{l}{$^{d}$ Number of points used to determine the maximum.} \\
\end{tabular}
\end{center}
\end{table}

\subsection{Light Variation during Mini-Outbursts}

\begin{figure}
  \begin{center}
    \FigureFile(88mm,110mm){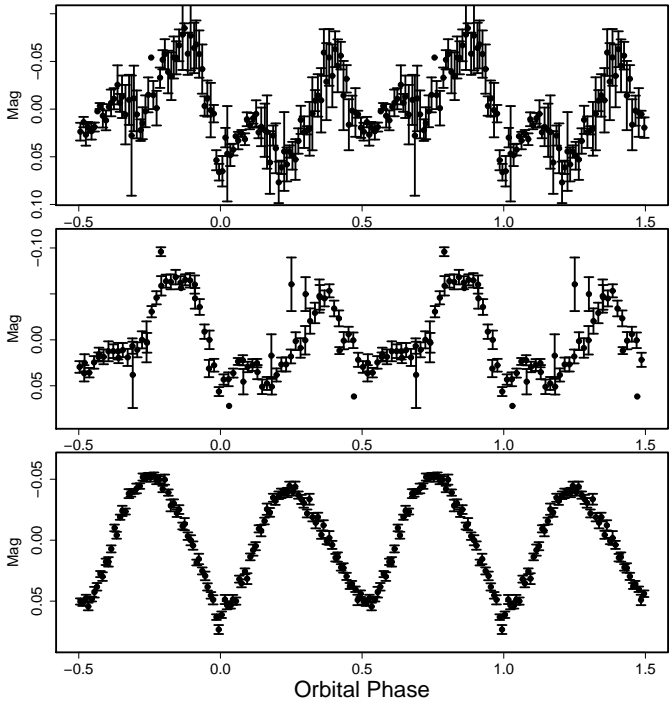}
  \end{center}
  \caption{Phase-averaged light curves during mini-outbursts.
     (Upper): the first mini-outburst
     (Middle): the second mini-outburst
     (Lower): quiescence.}
  \label{fig:j0804quiph}
\end{figure}

   We analyzed the ``mini-outbursts'' reported in \citet{zha08j0804}.
The phase-averaged light curves during the two mini-outbursts have
similar shape, and were distinct from that in quiescence
(figure \ref{fig:j0804quiph}).  The light curve during these
mini-outbursts somewhat resembles the orbital light curve during
the rebrightening and final fading phases.  The sharp eclipse feature
was absent during two mini-outbursts, suggesting that these brightenings
were not associated with an enhanced hot spot at a disk radius comparable
to that in outburst.

\begin{figure}
  \begin{center}
    \FigureFile(88mm,110mm){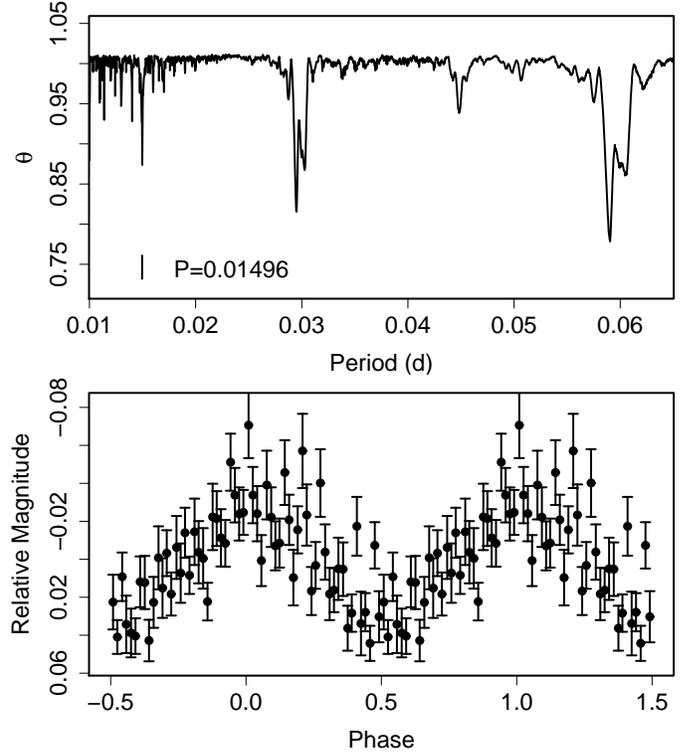}
  \end{center}
  \caption{Period analysis of the second mini-outburst after subtracting
     the quiescent orbital variation.
     (Upper): PDM analysis.  An excess component longer than the orbital
     period was present.  These excess variations were apparently
     caused by a variation with a period of 0.01496(3) d.
     (Lower): Mean profile of the 0.01496 d variation.}
  \label{fig:j0804minipdm}
\end{figure}

   A period analysis during the (better-sampled) second mini-outburst,
after subtracting the orbital light curve in quiescence.
These was an excess component longer than the orbital period,
as was seen in the rebrightening and final decline phases.
Although these variations can naturally be attributed to superhumps,
there were four peaks in one cycle.  The basic period was
0.01496(6) d, close to the one fourth of the superhump period
(figure \ref{fig:j0804minipdm}).  The period was different from
the 12.6-min periodicity supposed to arise from the pulsation of
the white dwarf \citep{pav08j0804WD}.  We regard that superhumps
were transiently excited during these mini-outbursts, although
it was insufficient to trigger a true outburst.  The brightness
increase may be understood as a result of increased tidal dissipation.

\subsection{SDSS J0804 as a WZ Sge-Type Dwarf Nova}

   Based on the presence of eleven rebrightenings, we classified the
superoutburst as type-B according to the classification scheme
by \citet{ima06tss0222}.
The apparent absence of flat quiescence between rebrightenings, short
($\sim$ 3 d) recurrent time of rebrightenings, and the relatively small
(1.5--2.0 mag) amplitudes of rebrightenings might place the superoutburst
intermediate between type-A and type-B.

   Although \citet{she07j0804} argued that the amplitude of the outburst
(up to 5 mag) is relatively small for a WZ Sge-type dwarf nova,
this is probably because the early stage of the superoutburst was missed.
If the evolution of the superoutburst was similar to that of V455 And,
another high-inclination WZ Sge-type dwarf nova, the true maximum
might have reached 11 magnitude.
The high orbital inclination probably partly contributes to the
low outburst amplitude.
The overall behavior of superhumps, however, and their $O-C$,
persistent long-period superhumps even after the rebrightening phase,
and the low $\epsilon$ all support the classification of this object
as a genuine WZ Sge-type object.  A detection of early superhumps during
future superoutbursts is awaited.

   Only two systems with type-B superoutbursts have been measured for
$\epsilon$: EG Cnc (0.6 \%) and AL Com (0.9 \%).  The $\epsilon$ (0.9 \%)
of SDSS J0804 is similar to those of previously known systems.

   Although no 2MASS counterpart of this object was detected, we performed
quiescent infrared photometry using OAO/ISLE \citep{ISLE}
in 2007 March, about one year after the superoutburst.
The infrared magnitudes were estimated to be $J$=17.29(0.05), $H$=16.97(0.05)
and $K_{\rm s}$=16.41(0.06), respectively.  The resultant color indices
(assuming mean $V$=17.1 at this epoch, \cite{zha08j0804}) $V-J = +0.2$
and $J-K_{\rm s} = +0.9$ are compatible with insignificant contribution
in the near-infrared as expected from the small $\epsilon$, and
consequently, a small mass of the secondary (cf. \cite{osa85SHexcess}).

   The presence of grazing eclipses is expected to provide a powerful
tool in diagnosing the variation of the disk radius and the luminosity
of the superhump light source or the hot spot.  Being the single known
eclipsing object exhibiting distinct multiple rebrightenings, detailed
observations of the next superoutburst of SDSS J0804 are expected to
provide a wealth of clue to understanding the origin of multiple
rebrightenings in WZ Sge-type dwarf novae.

\vskip 3mm

This work was supported by the Grant-in-Aid for the Global COE Program
``The Next Generation of Physics, Spun from Universality and Emergence''
from the Ministry of Education, Culture, Sports, Science and Technology
(MEXT) of Japan.
The work is partially supported (YS) by grants NSh-1685.2008.2,
RNP-2906, RFBR-08-02-01220, and by grant F 25.2/139 of FRSF (EP).
We acknowledge with thanks the variable star
observations from the AAVSO International Database contributed by
observers worldwide and used in this research.

\end{document}